\documentclass[prb,10pt,twocolumn]{revtex4}

\usepackage{epsfig}
\graphicspath{{/}}

\newcommand{\ul}[1]{\underline{#1}}

\begin{document}

\title{Light Absorption by Excitons and Trions in Monolayers of Metal Dichalcogenide MoS$_{2}$: Experiments and Theory}

\author{Changjian Zhang, Haining Wang, Weimin Chan, Christina Manolatou, Farhan Rana}
\affiliation{School of Electrical and Computer Engineering, Cornell University, Ithaca, NY 14853}
\email{fr37@cornell.edu}

\begin{abstract}
We measure the optical absorption spectra and optical conductivities of excitons and trions in monolayers of metal dichalcogenide MoS$_{2}$ and compare the results with theoretical models. Our results show that the Wannier-Mott model for excitons with modifications to account for small exciton radii and large exciton relative wavefunction spread in momentum space, phase space blocking due to Pauli exclusion in doped materials, and wavevector dependent dielectric constant gives results that agree well with experiments. The measured exciton optical absorption spectra are used to obtain experimental estimates for the exciton radii that fall in the $7-10\AA$ range and agree well with theory. The measured trion optical absorption spectra are used to obtain values for the trion radii that also agree well with theory. The measured values of the exciton and trion radii correspond to binding energies that are in good agreement with values obtained from first principles calculations.        
\end{abstract}
                                    
\maketitle

\section{Introduction}
Two dimensional (2D) metal dichalcogenides have emerged as important materials for a variety of different applications in electronics and optoelectronics \cite{fai10,fai12,fai13,xu13,wang10,kis11,kis13}. Particularly distinguishing features of 2D metal dichalcogenides are the large exciton and trion binding energies in these materials. The exciton and trion binding energies in 2D chalcogenides are almost an order of magnitude larger compared to other bulk semiconductors \cite{fai10,fai13,xu13}. The large exciton and trion binding energies imply that many body interactions play an important role in determining the electronic and optical properties of these materials. Energy bandstructures, including exciton wavefunctions and binding energies, and optical absorption spectra for these materials have been obtained using a variety of different theoretical techniques including ab-initio calculations \cite{Lam12,Louie1,timothy}. The question that remains to be answered is how well the calculated exciton and trion binding energies and wavefunctions agree with experiments and what essential physics has to be included in the models to obtain good quantitative agreements with measurements. The goal of this paper is to contribute to the answer of this question. 

In this work, we measure the absorption spectra of excitons and trions (in transmission configuration) in monolayers of a dichalcogenide (MoS$_{2}$) and use it to extract information about the exciton and trion radii, optical oscillator strengths, and binding energies. Our work shows that the traditional Wannier-Mott exciton model \cite{huagbook}, with a few modifications, is able to describe the exciton fairly well in 2D dichalcogenides. The modifications involve using accurate bandstructures of conduction and valence bands for large wavevectors, incorporating phase space blocking due to Pauli exclusion in doped samples, and using a wavevector dependent dielectric constant that takes into account the finite thickness of dichalcogenide monolayers. Accurate bandstructures are required for large wavevectors because exciton and trion radii in dichalcogenides are of the order of $\sim10\AA$ and, therefore, the exciton and trion relative wavefunctions in the wavevector (or momentum) space spread out to large wavevectors. The small exciton and trion radii also imply that just the bandedge optical matrix elements cannot be used to obtain accurate exciton and trion optical absorption spectra, and matrix elements for large wavevectors are required. We show that with these modifications the measured exciton and trion optical absorption spectra (and optical conductivity values) agree well with the theory. Measurment of the optical absorption spectra and oscillator strengths can be used to estimate the exciton and trion radii and, to the best of our knowledge, this paper reports experimental estimates for the exciton and trion radii in a dichalcogenide for the first time. Our work also shows that the exciton and trion binding energies that follow from the experimentally determined exciton and trion radii agree well with the results obtained previously from first principles calculations. The results presented in this paper are expected to contribute to the field of metal dichalcogenide optoelectronics.      

Section \ref{sec:bands} discusses the basic bandstructure and optical properties of metal dichalcogenides. Section \ref{sec:excitons} discusses the theoretical model for excitons, derives expressions for the exciton optical conductivity and absorption spectra, presents experimental results and compares them to the theory. Section \ref{sec:trions} discusses the theoretical model for trions and compares the experimental results to the theory.

\section{Bandstructure and Optical Properties} \label{sec:bands}
\begin{figure}[bp]
  \begin{center}
   \epsfig{file=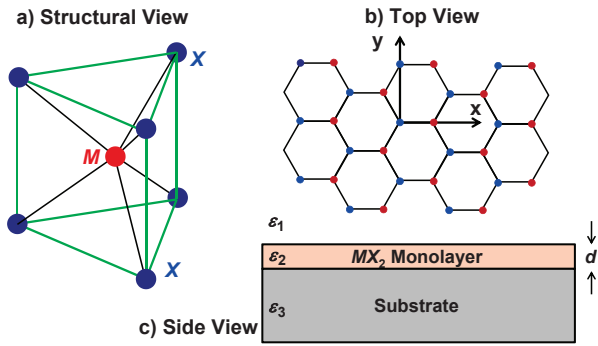,angle=0,width=0.45\textwidth}
      \caption{(a) A unit cell of $MX_{2}$. (b) $MX_{2}$ monolayer top view. (c) A $MX_{2}$ monolayer sandwiched between two dielectrics.}
    \label{figMX3}
  \end{center}
\end{figure}

The crystal structure of a monolayer of group-VI dichalcogenides $MX_{2}$ (e.g. $M$=Mo,W and $X$=S,Se) consist of $X$-$M$-$X$ layers, and within each layer the $M$ atoms (or the $X$ atoms) form a 2D hexagonal lattice. Each $M$ atoms is surrounded by 6 nearest neighbour $X$ atoms in a trigonal prismatic geometry with $D_{3h}^{1}$ symmetry, as shown in Fig.~\ref{figMX3}. The valence band maxima and conduction band minima occur at the $K$ and $K'$ points in the Brillouin zone. Symmetry can dictate the form of the Bloch functions near the $K$($K'$) points. Calculations based on first principles have shown that near the band extrema most of the weight in the Bloch states resides on the d-orbitals of $M$ atoms \cite{groot87,mattheiss73,wold87,Lam12,Louie1,Falko13,yao12,timothy}. Assuming only d-orbitals for the conduction and valence band states, and including spin-orbit coupling, one obtains the following simple spin-dependent tight-binding hamiltonian (in matrix form) near the $K$($K'$) points \cite{yao12},
\begin{equation}
\left[
\begin{array}{cc}
\Delta/2 & \hbar v k_{-} \\
\hbar v k_{+} & -\Delta/2 + \lambda \tau \sigma
\end{array} \right]    \label{eq:H1}
\end{equation}
Here, $\Delta$ is related to the material bandgap, $\sigma=\pm1$ stands for the electron spin, $\tau=\pm1$ stands for the $K$ and $K'$ valleys, $2\lambda$ is the splitting of the valence band due to spin-orbit coupling, $k_{\pm}=\tau k_{x}\pm ik_{y}$, and the velocity parameter $v$ is related to the coupling between the orbitals on neighbouring $M$ atoms. From density functional theories \cite{Lam12,Falko13}, $v\approx 5-6 \times 10^5 $ m/s. The wavevectors are measured from the $K$($K'$) points. The momentum matrix element between the conduction and valence band states near $K$($K'$) points follows from the above Hamiltonian,
\begin{equation}
\vec{P}_{vc}(\vec{k}\approx 0) = m_{o}v(\tau\hat{x} + i\hat{y}) \label{eq:matrix}
\end{equation}
Here, $m_{o}$ is the free electron mass.
The Hamiltonian and the optical (momentum) matrix element given above are accurate only near the band edges. Later in this paper, we will need to modify the Hamiltonian and obtain results that are accurate for large wavevectors.

\section{Absorption of Light by Excitons} \label{sec:excitons}

\subsection{Exciton States}
We assume that the initial state $|\psi_{i}\rangle$ of the semiconductor consists of a completely filled valence band and a conduction band with an electron density $n_{e}$ distributed according to the Fermi-Dirac distribution $f_{c}(\vec{k})$. The initial state thus belongs to a thermal ensemble and the (average) energy of the ground state is $E_{i}$. Without losing generality, we restrict ourselves to the valley $\tau=1$ where the top most valence band is occupied by spin-up ($\sigma=1$) electrons. Only excitons with zero in-plane momentum are created by normally incident radiation. An exciton state with zero in-plane momentum can be constructed from the initial state as follows,
\begin{equation}
|\psi_{ex}\rangle = \frac{1}{\sqrt{A}}\sum_{\vec{k}} \frac{\phi(\vec{k})}{N_{ex}(\vec{k})} c_{\vec{k}, \uparrow}^{\dagger} b_{\vec{k}, \uparrow}|\psi_{i} \rangle \label{eq:exciton}
\end{equation} 
In the above Equation, $c_{\vec{k},\uparrow}$ and $b_{\vec{k},\uparrow}$ are the destruction operators for the spin-up conduction and valence band states, respectively, with momentum $\vec{k}$. $A$ is the area of the monolayer. The above state is that of a Wannier exciton and we are assuming that the Wannier exciton theory is valid for $MX_{2}$ materials. The normalization factor $N_{ex}(\vec{k})$, which equals $\sqrt{1 - f_{c}(\vec{k})}$, will prove useful later. The exciton state is normalized such that $\left\{ \langle \psi_{ex}|\psi_{ex} \rangle\right\}_{th} =1$, where the curly brackets represent averaging with respect to the thermal ensemble. This normalization gives, 
\begin{equation}
\int \frac{d^{2}\vec{k}}{(2\pi)^{2}} |\phi(\vec{k})|^{2} = 1 
\end{equation}
The state in Eq.~\ref{eq:exciton} is an exact eigenstate of the interacting Hamiltonian only when the electron density $n_{e}$ is zero. Since $n_{e} \ne 0$, we assume that the state in Eq. ~\ref{eq:exciton} is a variational state and the parameter $\phi(\vec{k})$ can be varied to minimize the expectation value of the energy, $\left\{ \langle \psi_{ex}| \hat{H} | \psi_{ex} \rangle \right\}_{th}$, subject to the normalization constrain. This results in the following eigenvalue equation that resembles the traditional Beth-Saltpeter equation for excitons \cite{huagbook},
\begin{eqnarray}
& & \left[\bar{E}_{c}(\vec{k}) - \bar{E}_{v}(\vec{k}) \right]\phi(\vec{k}) \nonumber \\
& & - \frac{\sqrt{1-f_{c}(\vec{k})}}{A} \sum_{\vec{q}} V(\vec{q}) \phi(\vec{k}-\vec{q}) \sqrt{1-f_{c}(\vec{k}-\vec{q})} \nonumber \\
& & = (E_{ex}-E_{i})\phi(\vec{k}) \label{eq:eigen}
\end{eqnarray}
Here, $E_{ex}$ is the (average) energy of the state $| \psi_{ex} \rangle$ and $E_{ex}-E_{i} = E_{g} - E_{exb}$, where $E_{exb}$ is the exciton binding energy. $V(\vec{q})$ is the 2D Coulomb potential and equals $e^{2}/2\epsilon_{o} \epsilon(\vec{q})q$, and $\bar{E}_{c}(\vec{k})$ and $\bar{E}_{v}(\vec{k})$ are the conduction and valence band dispersions, respectively, including exchange corrections \cite{huagbook}. The same eigenvalue equation is obtained using Green's functions and summing the ladder diagrams under the assumption that $n_{e} \ne 0$ \cite{mahan}. The choice of the appropriate dielectric constant $\epsilon(\vec{q})$ is discussed in Appendix A. The eigenvalue equation obtained above is Hermitian and the eigenfunctions are orthogonal in the sense,
\begin{equation}
\int \frac{d^{2}\vec{k}}{(2\pi)^{2}} \phi^{*}_{m}(\vec{k}) \phi_{p}(\vec{k}) = \delta_{m,p} 
\end{equation}
Obtaining a Hermitian eigenvalue equation was the motivation in using the particular normalization factor $N_{ex}(\vec{k})$ in Eq.~\ref{eq:exciton}. Solutions of the above equation represent bound electron-hole pairs (excitons) as well as electron-hole scattering states. The latter are excluded from the discussion that follows since their inclusion leads to modification in the absorption spectrum near the band edge far from the fundamental exciton line \cite{huagbook,runge01}. Since the eigenvalue equation is Hermitian, a variational approach can be used to obtain an approximate solution for the lowest exciton state. Following Schmitt-Rink et.~al.~\cite{schmittrink1}, we assume the following variational solution,
\begin{equation}
\phi(\vec{k}) = \frac{\sqrt{8\pi} a}{(1 + (ka)^2)^{3/2}} \, \, \, \rightarrow \, \, \, \phi(\vec{r}) = \sqrt{\frac{2}{\pi}} \frac{1}{a}e^{-r/a} \label{eq:orbital}
\end{equation} 
The solution corresponds to the exact exciton wavefunction when $n_{e}=0$ and screening is local ($\epsilon(\vec{q})$ is independent of $\vec{q}$). The radius parameter $a$ can be varied to estimate the eigenvalue $E_{ex}-E_{i}$.  

\subsection{Exciton Optical Conductivity}
We assume light of frequency $\omega$ and intensity $I_{o}$ (units: W-cm$^{-2}$) incident normally on the $MX_{2}$ monolayer. We assume linearly polarized light. In the plane of the monolayer, the vector potential describing the incident light is $\hat{n} A_{o} cos(\omega t)$, where $\hat{n}$ is the polarization vector. It follows that $I_{o} = \omega^{2}A_{o}^{2}/2\eta_{o}$, where $\eta_{o}$ is the free space impedance. The interaction between the spin-up electrons in the valley $\tau=1$ and light is given by the time-dependent Hamiltonian,
\begin{eqnarray}
H_{int}(t) & = & H_{+} e^{-i\omega t} +  H_{-} e^{i\omega t} \nonumber \\
& = & \frac{eA_{o}}{2m_{o}} \sum_{\vec{k}}  \vec{P}_{cv}(\vec{k}).\hat{n} e^{-i\omega t} c_{\vec{k}, \uparrow}^{\dagger} b_{\vec{k}, \uparrow}  + h.c. \label{eq:light_H}
\end{eqnarray}
The rate $R_{ex}$ (units: s$^{-1}$-cm$^{2}$) at which excitons are generated by the absorption of light can be found using Fermi's golden rule assuming a finite broadening,
\begin{eqnarray}
R_{ex} & = & \frac{2\pi}{\hbar} \frac{1}{A} \left\{ \left| \langle \psi_{ex} | H_{+} | \psi_{i} \rangle \right|^{2} \right\}_{th} \nonumber \\
& & \times \frac{\Gamma_{ex}/\pi}{\left( E_{ex} - E_{i} - \hbar \omega \right)^{2} + \Gamma_{ex}^{2}} \nonumber \\
& = & \frac{2\pi}{\hbar} \left( \frac{eA_{o}}{2m_{o}} \right)^{2}  \left|\chi_{ex}(\vec{r}=0)\right|^{2} \nonumber \\
& & \times \frac{\Gamma_{ex}/\pi}{\left( E_{ex} - E_{i} - \hbar \omega \right)^{2} + \Gamma_{ex}^{2}}
\end{eqnarray}
Here,
\begin{equation}
\chi_{ex}(\vec{r}) = \int \frac{d^{2}\vec{k}}{(2\pi)^{2}} \vec{P}_{cv}(\vec{k}).\hat{n} \, \, \phi(\vec{k}) \sqrt{1-f_{c}(\vec{k})} e^{i\vec{k}.\vec{r}}
\end{equation}
$\chi_{ex}(\vec{r})$ incorporates the reduction in the exciton oscillator strength due to Pauli-blocking. Effects due to Pauli-blocking come about due to the presence of the factor $\sqrt{1-f_{c}(\vec{k})}$ in the expression above and also from the modification of the function $\phi(\vec{k})$. The total energy absorption rate from both $K$ and $K'$ valleys, $2\hbar\omega R_{ex}$, can be written in terms of the exciton contribution $\sigma_{ex}(\omega)$ to the optical conductivity as, $\Re\{\sigma_{ex}(\omega)\}\eta_{o} I_{o}$, where, 
\begin{eqnarray}
\Re\{\sigma_{ex}(\omega)\} & = &  \frac{e^{2}}{4\hbar}\left\{ \frac{8\hbar}{m_{o}^{2}\omega} \left|\chi(\vec{r}=0)\right|^{2} \right.  \nonumber \\
& & \left. \times \frac{\Gamma_{ex}}{\left( E_{ex}-E_{i} - \hbar \omega \right)^{2} + \Gamma_{ex}^{2}} \right\} \label{eq:excond1}
\end{eqnarray}
Scattering processes as well as inhomogeneous broadening are both expected to contribute to the absorption width $\Gamma_{ex}$ \cite{excitonbook}.  

\subsection{Experimental Results and Discussion: Exciton Absorption in MoS$_{2}$ Monolayers} \label{sec:expex}
\begin{figure}
  \begin{center}
   \epsfig{file=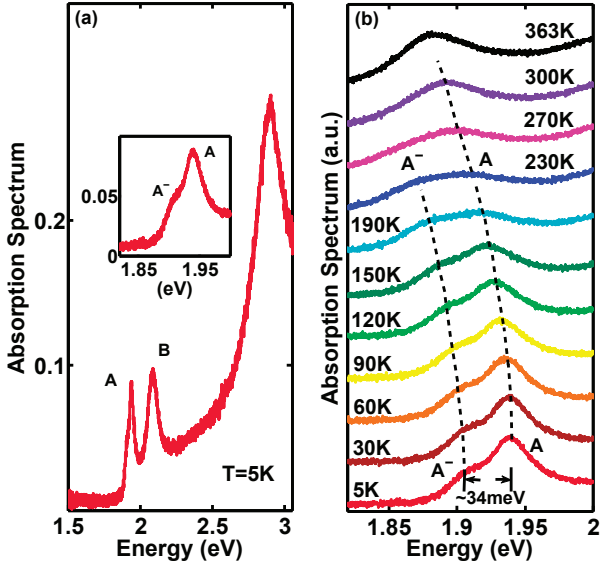,angle=0,width=0.48\textwidth}
    \caption{(a) Measured broadband absorption spectrum $A(\omega)$ of a MoS$_2$ monolayer at $T$=5K is plotted. Two main absorption resonances, $A$ and $B$, attributed to excitons are visible together with a smaller $A^{-}$ attributed to trions. The inset shows the $A$ and $A^-$ peaks in greater detail. (b) Measured absorption spectrum of a MoS$_2$ monolayer near the $A$ and $A^{-}$ peaks is plotted for different temperatures. The curves for different temperatures are given offsets in the vertical direction for clarity. The $A^-$-trion absorption peak is clearly visible and distinguishable from the $A$-exciton absorption peak for all temperatures below 230K. The dashed lines are guides to eyes showing the evolution of the exciton ($A$) and trion ($A^-$) peaks as a function of temperature.}
    \label{dpeakdtrans}
  \end{center}
\end{figure}

Optical absorption experiments on MoS$_{2}$ monolayer flakes were performed at different temperatures in the transmission configuration. The  MoS$_{2}$ flakes were exfoliated from bulk MoS$_{2}$ crystals and transferred onto quartz substrates. Typical flake sizes were $10-15$ $\mu m$. All experiments were performed using a confocal microscope setup with a 100X objective. MoS$_2$ monolayer samples were identified by inspection under an optical microscope and confirmed by Raman spectroscopy \cite{lee10}. The samples were annealed in vacuum at 360K for 8 hours. The samples were found to be n-doped. The electron density was estimated from Raman measurements to be in the 2-4$\times10^{12}$ 1/cm$^{2}$ range \cite{sood12}. Electrical measurements on similar samples on oxide-coated doped silicon substrates (with electrostatic gating) yielded intrinsic electron densities in the same range. The samples were placed in a helium-flow cryostat and measurements were done in the temperature range 5K-363K. Sample transmission spectrum was obtained by dividing the spectrum $T(\omega)$of the transmitted light through the sample and the quartz substrate by the reference spectrum $T_{ref}(\omega)$ obtained for just the quartz substrate. Absorption spectrum $A(\omega)$ was calculated by subtracting the sample transmission spectrum from unity. A typical broadband absorption spectrum of MoS$_2$ monolayers obtained at 5K is shown in Fig.~\ref{dpeakdtrans}(a). Two large exciton resonances, labelled $A$ and $B$, are visible at $\sim1.9$ eV and $\sim2.1$ eV, respectively\cite{fai10,fai13,Evans65}. These are attributed to excitons involving the conduction band and the two valence bands (which are split due to spin-orbit coupling) near the $K$($K'$) points. A smaller absorption peak, labelled $A^{-}$, is also visible near the $A$ peak and is attributed to trion absorption \cite{fai13}. Fig.~\ref{dpeakdtrans}(b) shows absorption spectra for a MoS$_2$ monolayer near the $A$ and $A^{-}$ peaks for different temperatures in the 5K-363K temperature range. Distinct $A$ and $A^{-}$ absorption peaks are visible at all temperatures below 230K. Both the exciton and trion peak positions red-shift as the temperature increases. We attribute this red-shift to the temperature dependence of the material bandgap \cite{Var67}. The $A$ and $A^{-}$ peaks broaden as the temperature increases and become almost indistinguishable above 230K. The temperature-dependent broadening is attributed to increase in the scattering rates with temperature. The splitting in the exciton and trion absorption peaks is $\sim$34 meV at 5K and does not change much as the temperature is increased to 230K. Since exciton and trion peak splitting is a function of the carrier density \cite{fai13}(also see the discussion in Sec.\ref{subsec:trion_abs}), the electron density in our samples does not change significantly in the 5K-230K temperature range and this was confirmed using measurements described above \cite{sood12}.       
\begin{figure}[tbp]
  \begin{center}
   \epsfig{file=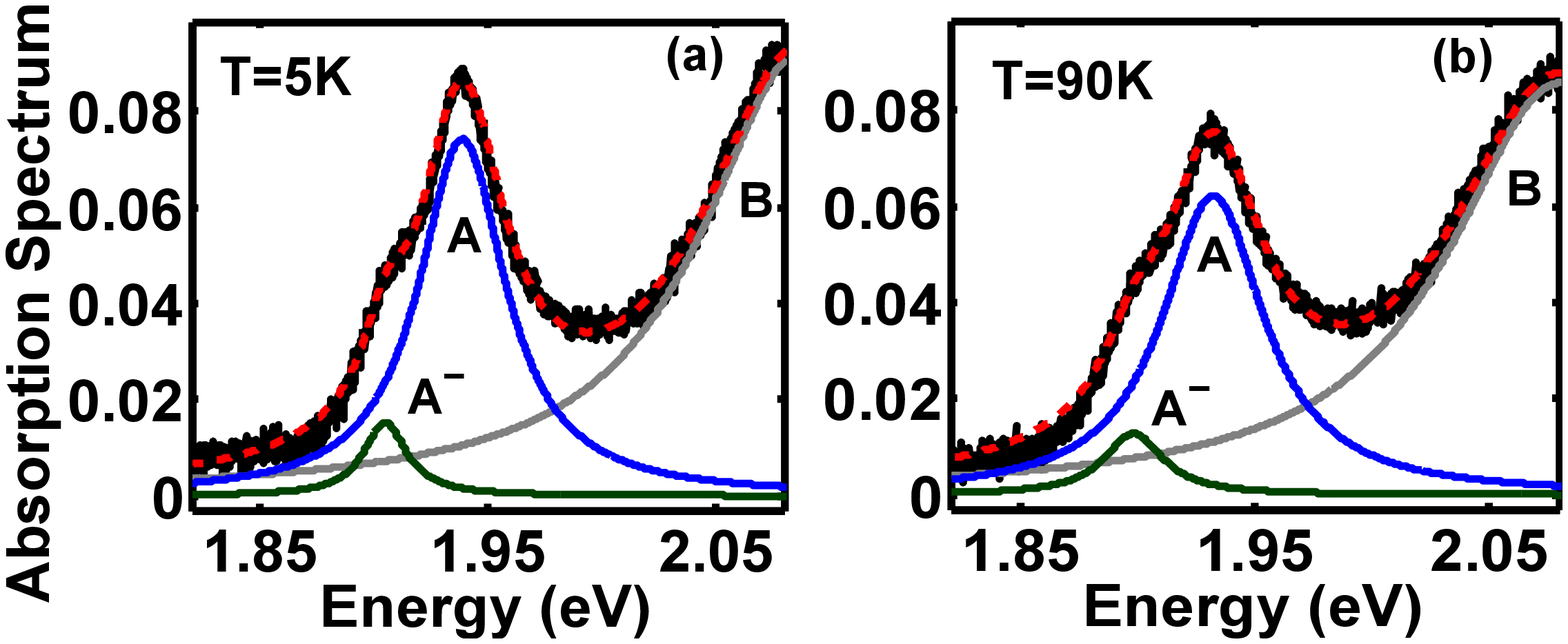,angle=0,width=0.48\textwidth}
    \caption{Extracted contributions of $A$-excitons (blue) and $A^-$-trions (green) to the measured absorption spectra (black) at 5K (a) and at 90K (b) are shown together with the data (black) and the quality of the fit (red-dashed). The extracted contribution of $B$-excitons (gray) is also visible in the Figure.}
    \label{triple_fitting}
  \end{center}
\end{figure} 
\begin{figure}[tbp]
  \begin{center}
   \epsfig{file=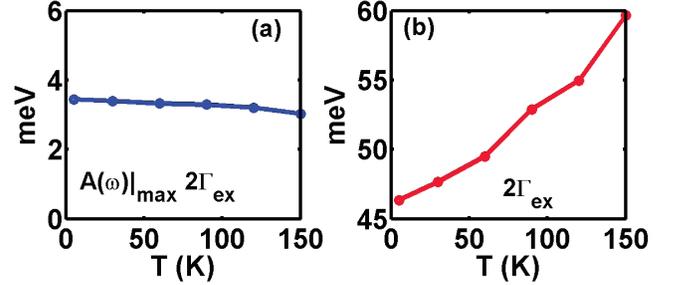,angle=0,width=0.48\textwidth}
    \caption{(a) The product of the peak absorption $A(\omega)|_{max}$ and the width $2\Gamma_{ex}$ of the extracted $A$-exciton absorption spectrum is plotted as a function of the temperature. (b) The full width $2\Gamma_{ex}$ of the extracted $A$-exciton absorption spectrum is plotted as a function of temperature.}
    \label{ex_gamma}
  \end{center}
\end{figure} 
The measured absorption spectrum for normal incident light can be expressed in terms of the optical conductivity,
\begin{equation}
A(\omega) = 1 - \frac {T(\omega)}{T_{ref}}\approx \frac{2Re\left\{\sigma(\omega)\right\}\eta_o}{1+n_{sub}} \label{eq:exspectra}
\end{equation}
where, $n_{sub}$ is the refractive index of the quartz substrate. In general, the optical conductivity has contributions from excitons, trions, and from other correlated electron-hole states~\cite{huagbook}. In this Section, we will focus on the exciton absorption resonance ($A$ peak). Eq.~(\ref{eq:excond1}) for the exciton optical conductivity and Eq.~(\ref{eq:trioncond1}) for the trion optical conductivity (discussed later in this paper) were used to fit the measured absorption spectra and extract the contributions of excitons and trions. Fig.~\ref{triple_fitting}(a) shows the extracted contributions from $A$-excitons and $A^-$-trions along with the measured absorption data at 5K and 90K. The Figure shows that the theoretical spectra fit the data well, and that individual contributions from excitons and trions can be reliably extracted from this fitting procedure. Fig.~\ref{ex_gamma} shows that the product of the peak absorption $A(\omega)|_{max}$ and the width $2\Gamma_{ex}$ of the extracted $A$-exciton absorption spectrum varies little with temperature in the 5K-150K range (there is in fact a $\sim$11$\%$ decrease going from 5K to 150K). In the same temperature range, $2\Gamma_{ex}$ varies from 45 meV at 5K to 60 meV at 150K (Fig.~\ref{ex_gamma}(b)).  
\subsubsection{Exciton Radius}
Ignoring Pauli-blocking and using the simple wavevector-independent expression for the momentum matrix element given in Eq.~(\ref{eq:matrix}), the exciton optical conductivity and the product $A(\omega)|_{max} 2\Gamma_{ex}$ are found to be,
\begin{equation}
\Re\{\sigma_{ex}(\omega)\} = \frac{2e^{2}v^{2}}{\omega}\left( \frac{2}{\pi a^{2}} \right) \frac{\Gamma_{ex}}{\left( E_{ex}-E_{i} - \hbar \omega \right)^{2} + \Gamma_{ex}^{2}} 
\end{equation}
\begin{equation}
A(\omega)|_{max} 2\Gamma_{ex} = \frac{16\eta_o}{1+n_{sub}} \frac{e^2 v^2}{\pi \omega} \left(\frac{1}{a^{2}}\right) \label{eq:product}
\end{equation} 
It is tempting to use the result in Eq.~(\ref{eq:product}) to extract the value of the exciton radius $a$. Using the data shown in Fig.~\ref{ex_gamma}(a), the exciton radius is found to be $\sim 16.8\AA$. The small exciton radius implies that the exciton wavefunction occupies an area in momentum space of radius at least a few nm$^{-1}$ and, therefore, the use of a wavevector-independent expression for the momentum matrix element is not justified. In addition, the reduction in the phase space due to Pauli-blocking cannot be ignored in the case of doped samples such as the ones used in this work. Accurate energy dispersions and momentum matrix elements are needed for large wavevectors (at least a few nm$^{-1}$) in order to accurately describe exciton oscillator strengths. Following Kormanyos et.~al.~\cite{Falko13}, we use Lowdin approximation to a 4-band model. This procedure adds the following matrix to the Hamiltonian given earlier in (\ref{eq:H1}), 
\begin{equation}
\left[
\begin{array}{cc}
\alpha k^{2} & \kappa k_{+}^2 - \frac{\eta}{2}k^{2}k_{-}    \\
\kappa k_{-}^{2} - \frac{\eta}{2}k^{2}k_{+} & \beta k^{2}
\end{array} \right]    \label{eq:H2}
\end{equation}
\begin{figure}[tbp]
  \begin{center}
   \epsfig{file=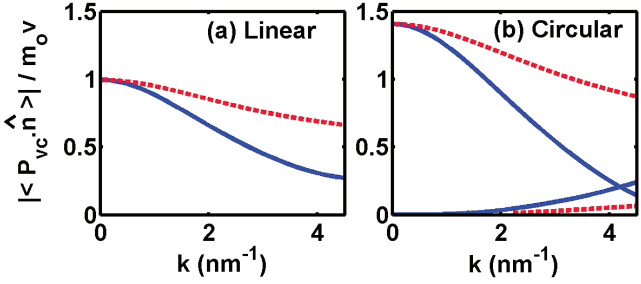,angle=0,width=0.48\textwidth}
    \caption{The angle-averaged interband momentum matrix elements $\langle \vec{P}_{vc}.\hat{n} \rangle$, normalized to $m_{o}v$, for linear and the two circular polarizations are plotted as a function of the wavevector (measured from the $K(K')$ points) for the case of the simple Hamiltonian given in (\ref{eq:H1}) (dashed) and including the corrections given in (\ref{eq:H2}) (solid).} 
    \label{mom_matrix}
  \end{center}
\end{figure} 
\begin{figure}[tbp]
  \begin{center}
   \epsfig{file=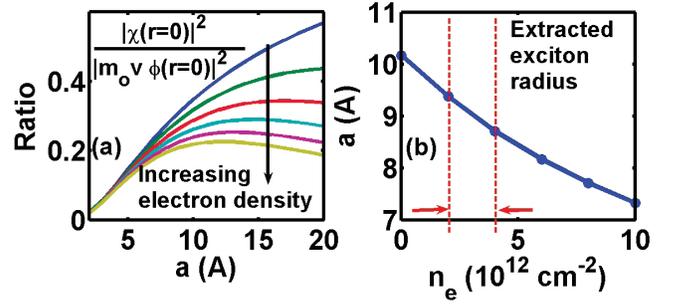,angle=0,width=0.48\textwidth}
    \caption{(a) The ratio of the exciton oscillator strengths with and without the effects of the wavevector dependent momentum matrix element and Pauli blocking is plotted as a function of the exciton radius for different electron densities ($n_{e}=0,2,4,6,8,10 \times 10^{12}$ cm$^{-2}$). T=5K. (b) The exciton radius extracted from the measurements is plotted as a function of assumed values for the electron density. The dashed lines indicate the range for the measured electron densities in our samples.} 
    \label{ex_radius}
  \end{center}
\end{figure}
The values of the parameters $\alpha$, $\beta$, $\kappa$ and $\eta$ that best fit density functional theory (DFT) results are 1.72 eV$\AA^{2}$, -0.13  eV$\AA^{2}$, -1.02  eV$\AA^{2}$, and 8.52  eV$\AA^{3}$, respectively \cite{Falko13}. These values results in $m_{e}\sim 0.5 m_{o}$ and $m_{h}\sim 0.6 m_{o}$ near the band edges. The resulting angle-averaged interband momentum matrix elements, normalized to $m_{o}v$, for linearly and circularly polarized light are plotted as a function of the wavevector in Fig.~\ref{mom_matrix} and show that the reduction in the matrix elements with the wavevector are significant. Fig.~\ref{ex_radius}(a) plots the ratio of the exciton oscillator strengths with and without the effects of the wavevector dependent momentum matrix element and Pauli-blocking as a function of the exciton radius for different electron densities ($n_{e}=0,2,4,6,8,10 \times 10^{12}$ cm$^{-2}$ and T=5K). When the radius is very small, the oscillator strength is small because the exciton wavefunction spreads more in the momentum space and the matrix element is smaller for larger momenta. When the radius is very large, the oscillator strength is again small because the exciton wavefunction is localized near the $K(K')$ points in momentum space where, in doped samples, Pauli blocking is larger. Taking into account the wavevector dependence of the momentum matrix elements and Pauli-blocking, one obtains,
\begin{equation}
A(\omega)|_{max} 2\Gamma_{ex} = \frac{16\eta_o}{1+n_{sub}} \frac{e^2}{2 m_{o}^{2} \omega} |\chi(\vec{r}=0)|^{2} \label{eq:product2}
\end{equation} 
The above equation can be used to extract the exciton radius more reliably, provided the electron density is known. Fig.~\ref{ex_radius}(b) shows the extracted values of the exciton radius from the absorption data (Fig.~\ref{ex_gamma}(a)) for different assumed values of the electron density. The dashed lines in the Figure indicate the range for the measured electron densities in our samples (Section \ref{sec:expex}). The extracted values of the exciton radius come out to be in the 9.3-8.5$\AA$ range for carrier densities in the 2-4$\times10^{12}$ 1/cm$^{2}$ range, in excellent agreement with theoretical estimates based on first-principles approaches \cite{timothy,Louie1}. Note that the values of the product of the Fermi wavevector and the exciton radius $k_{F}a$ for these radii and densities are in the 0.23-0.30 range, indicating a small, but not insignificant, phase space filling in our doped samples. Also as a result of Pauli-blocking, our model predicts a reduction in the exciton oscillator strength by 10.5-12.5$\%$ with the increase in temperature from 5K to 150K for the same carrier densities, also in very good agreement with the data shown in Fig.~\ref{ex_gamma}(a).    
\begin{figure}[tbp]
  \begin{center}
   \epsfig{file=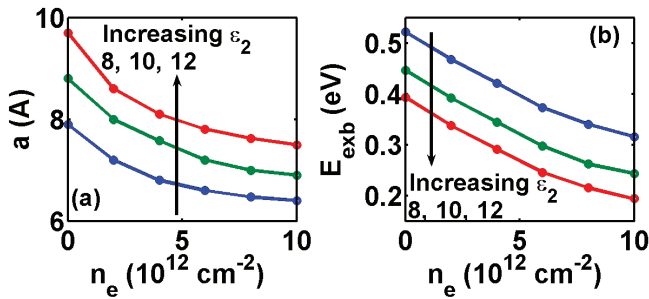,angle=0,width=0.48\textwidth}
    \caption{(a) The calculated exciton radius is plotted as a function of the electron density for different values of the MoS$_{2}$ dielectric constant $\epsilon_{2}$. T=5K. (b) The calculated exciton binding energy is plotted as a function of the electron density for different values of the MoS$_{2}$ dielectric constant $\epsilon_{2}$. T=5K.} 
    \label{ex_energy}
  \end{center}
\end{figure}
\subsubsection{Exciton Binding Energy} \label{sec:ex_energy}
To find the binding energy of the excitons, we need to solve Eq.~(\ref{eq:eigen}) using the variational solution in Eq.~(\ref{eq:orbital}). Since the radius of the excitons has already been estimated from light absorption measurements, the only parameter that is available to fit the data is the dielectric constant. 

We assume a MoS$_{2}$ monolayer of thickness $d$ and effective dielectric constant $\epsilon_2$ sandwiched between materials with dielectric constants $\epsilon_1$ and $\epsilon_3$, as shown in Fig.~\ref{figMX3}(c). For this geometry, the dielectric constant $\epsilon(\vec{q})$, which describes the Coulomb interaction between electrons and holes, is wavevector dependent~\cite{keldysh79} (see Appendix A for details). In our work, $\epsilon_3$ is 4.0 (quartz substrate) and $\epsilon_1$ is 1.0 (free space). The value of $\epsilon_2$ is used as an adjustable parameter. In a MoS$_{2}$ monolayer, the distance between the top and bottom sulphur atoms is $\sim 3.17\AA$. We assume an effective monolayer thickness of $6\AA$ ($d\approx6\AA$). Fig.~\ref{ex_energy} shows the values of the exciton radius and the exciton binding energy calculated using Eq.~(\ref{eq:eigen}) for different electron densities and for different values of the MoS$_{2}$ layer dielectric constant $\epsilon_{2}$. Note that a rather large value of $\epsilon_{2}$ ($\sim$12) is needed to match the calculated exciton radii to the measured values, $9.3-8.5 \AA$ (for electron densities 2-4$\times10^{12}$ 1/cm$^{2}$). Interestingly, this extracted value of $\epsilon_{2}$ matches well with the theoretical estimates for the bulk MoS$_{2}$ dielectric constant (see Appendix A for details). Following the work of Berkelbach et.~al.~\cite{timothy} and  Cudazzo et.~al.~\cite{cudazzo11}, one can find the value of the screening length parameter $r_{o}$, which is related to the in-plane polarizability of the MoS$_{2}$ monolayer, from the extracted value of the dielectric constant $\epsilon_{2}$. The resulting value of $r_{o}$ is $\sim36\AA$ and is in excellent agreement with first principles calculations of Berkelbach et.~.al~\cite{timothy} who obtained values for $r_{o}$ in the $30-40\AA$ range for a MoS$_{2}$ monolayer using different ab-initio techniques (see Appendix A for details). 

The values of the exciton binding energy that correspond to the extracted values of the exciton radius are in the 0.28-0.33 eV range.  Fig.~\ref{ex_energy} shows that a non-zero electron density can significantly modify the exciton radius and binding energy and that knowledge of the carrier density is necessary when comparing theory with experiments.  The decrease in the exciton radius with the electron density given by theory can be explained as follows. The variational approach used in this work minimizes the energy. When the electron density increases then, as a result of Pauli-blocking, the phase space available to the electron, which is interacting with the hole via the Coulomb potential, is reduced for small momenta. Energy can be minimized better if the exciton wavefunction spreads to larger momenta that are unoccupied. Consequently, the exciton radius decreases in real space.  The price paid in this trade-off is that the kinetic energy associated with the relative electron-hole motion increases when the exciton radius decreases and, therefore, the exciton binding energy also decreases. The calculated values of the exciton binding energy in the limit of zero electron density agree well with the values calculated theoretically from first-principles techniques \cite{Louie1,timothy}. Next, we discuss the trion absorption spectra.

\section{Absorption of Light by Trions} \label{sec:trions}

\subsection{Singlet Trion States}
A trion state can be a spin singlet or a triplet. Bound states for triplet trions have not been observed in two dimensions in either simulations \cite{suri01a} (for the case $m_{e} \approx m_{h}$ relevant for $MX_{2}$ monolayers) or in experiments in the absence of a magnetic field \cite{pepper02}, and therefore in this paper we consider only singlet trions. A trion is formed when a photoexcited electron-hole pair binds with an electron (or a hole) to form a negatively (or a positively) charged complex. Without losing generality, we restrict ourselves to negatively charged trions (relevant to n-doped samples). We also restrict ourselves to the case $\tau=1$ where the top most valence band is occupied by spin-up ($\sigma=1$) electrons.  We define the trion mass as $m_{tr} = 2m_{e} + m_{h}$ and the exciton mass as $m_{ex} = m_{e} + m_{h}$. As before, we assume that the initial state $|\psi_{i}\rangle$ of the semiconductor consists of a completely filled valence band and a conduction band with an electron density $n_{e}$ distributed according to the Fermi-Dirac distribution. A singlet trion state with momentum $\vec{Q}$ can be constructed from this initial state as follows,
\begin{eqnarray}
|\psi_{tr}(\vec{Q})\rangle  & = &  \frac{1}{A}\sum_{\vec{k}_{1},\vec{k}_{2}} \frac{\phi(\vec{k}_{1},\vec{k}_{2}, \vec{Q})}{N_{tr}(\vec{k_{1}},\vec{k_{2}},\vec{Q})}  \times \nonumber \\
&& c_{\ul{\vec{k}_{1}}, \downarrow}^{\dagger}  c_{\ul{\vec{k}_{2}}, \uparrow}^{\dagger} b_{\vec{k}_{1} + \vec{k}_{2} - \frac{m_{h}}{m_{tr}}\vec{Q}, \uparrow} c_{\vec{Q},\downarrow} |\psi_{i} \rangle   \label{eq:trion}
\end{eqnarray}
Here, the line under a vector, $\ul{\vec{k}}$, stands for $\vec{k}+(m_{e}/m_{tr})\vec{Q}$. The function $\phi(\vec{k}_{1},\vec{k}_{2},\vec{Q})$ is symmetric in its first two arguments. The normalization factor $N_{tr}(\vec{k_{1}},\vec{k_{2}},\vec{Q})$ equals,
\begin{equation}
\sqrt{f_{c}(\vec{Q})(1 - f_{c}(\ul{\vec{k}_{2}})) (1 - f_{c}(\ul{\vec{k}_{1}}))}
\end{equation}
Since $n_{e} \ne 0$, we again assume that the state in Eq.~\ref{eq:trion} is a variational solution and $\phi(\vec{k}_{1},\vec{k}_{2},\vec{Q})$ can be varied to minimize the expectation value of the energy, $\left\{ \langle \psi_{tr}| \hat{H} | \psi_{tr} \rangle \right\}_{th}$, subject to the normalization constrain. This results in a Hermitian eigenvalue equation for the trion wavefunction $\phi(\vec{k}_{1},\vec{k}_{2},\vec{Q})$ and the trion energy $E_{tr}(\vec{Q})$,
\begin{eqnarray}
&& \left[ \bar{E}_{c}(\ul{\vec{k_{1}}}) + \bar{E}_{c}(\ul{\vec{k_{2}}}) - \bar{E}_{c}(\vec{Q}) \right. \nonumber \\
&& \left. - \bar{E}_{v}(\vec{k_{1}}+\vec{k_{2}}-\frac{m_{h}}{m_{tr}}\vec{Q}) \right]\phi(\vec{k}_{1},\vec{k}_{2},\vec{Q}) \nonumber \\
&&  + \frac{\sqrt{1 - f_{c}(\ul{\vec{k}_{1}})} \sqrt{1 - f_{c}(\ul{\vec{k}_{2}})}}{A} \sum_{\vec{q}} \left[ V(\vec{q}) \phi(\vec{k_{1}}-\vec{q},\vec{k_{2}}+\vec{q},\vec{Q}) \right. \nonumber \\
&& \left. \sqrt{1 - f_{c}(\ul{\vec{k}_{1}}-\vec{q})} \sqrt{1 - f_{c}(\ul{\vec{k}_{2}}+\vec{q})}  \right] \nonumber \\
&& - \frac{\sqrt{1 - f_{c}(\ul{\vec{k}_{1}})} }{A} \sum_{\vec{q}} \left[ V(\vec{q})\, \phi(\vec{k_{1}}-\vec{q},\vec{k_{2}},\vec{Q}) \right. \nonumber \\
&& \left. \sqrt{1 - f_{c}(\ul{\vec{k}_{1}}-\vec{q})} \right] - \frac{\sqrt{1 - f_{c}(\ul{\vec{k}_{2}})} }{A} \sum_{\vec{q}} \left[ V(\vec{q})\, \phi(\vec{k_{1}},\vec{k_{2}-\vec{q}},\vec{Q}) \right. \nonumber \\
&& \left. \sqrt{1 - f_{c}(\ul{\vec{k}_{2}}-\vec{q})} \right] = (E_{tr}(\vec{Q})-E_{i})\phi(\vec{k}_{1},\vec{k}_{2},\vec{Q})  \label{eq:eigen2}
\end{eqnarray}
Here,
\begin{equation}
E_{tr}(\vec{Q}) - E_{i} = E_{g} - E_{exb} - E_{trb} - \frac{\hbar^{2}Q^{2}}{2m_{e}} \left( \frac{m_{ex}}{m_{tr}} \right)
\end{equation}
where $E_{trb}$ is the trion binding energy. The trion wavefunctions are orthogonal and normalized such that,
\begin{equation}
\int \frac{d^{2}\vec{k_{1}}}{(2\pi)^{2}} \frac{d^{2}\vec{k_{2}}}{(2\pi)^{2}} \phi^{*}_{m}(\vec{k}_{1},\vec{k}_{2},\vec{Q}) \phi_{p}(\vec{k}_{1},\vec{k}_{2},\vec{Q}) = \delta_{m,p} \end{equation}
Solutions of the above equation represent bound trion states as well as exciton-electron scattering states that correspond to a free electron interacting with a bound electron-hole pair. The latter are excluded from the discussion below for the sake of simplicity. Since the trion eigenvalue equation is Hermitian, a variational approach can be used to obtain an approximate solution for the lowest trion state. We use the variational solution proposed by Suris et.~al.~\cite{suri01},
\begin{eqnarray}
&& \phi(\vec{k}_{1},\vec{k}_{2},\vec{Q}) \approx \frac{1}{\sqrt{2}}\left[ 1 + 16 \frac{b^{2} c^{2}}{(b + c)^4} \right]^{-1/2} \times  \nonumber \\
&& \left\{ \frac{\sqrt{8\pi}b}{(1 + (k_{1}b)^2)^{3/2}} \frac{\sqrt{8\pi} c}{(1 + (k_{2}c)^2)^{3/2}}  + \left( \begin{array}{c} \vec{k_{1}} \rightarrow \vec{k_{2}} \\ \vec{k_{2}} \rightarrow \vec{k_{1}} \end{array} \right) \right\} \label{eq:soltrion}
\end{eqnarray}
which in real space corresponds to,
\begin{eqnarray}
& & \phi(\vec{r}_{1},\vec{r}_{2},\vec{Q}) \approx  \frac{\sqrt{2}}{\pi b c} \left[ 1 + 16 \frac{b^{2} c^{2}}{(b + c)^4} \right]^{-1/2} \nonumber \\
& & \times \left( e^{-|\vec{r}_{1}|/b - |\vec{r}_{2}|/c} + \left( \begin{array}{c} \vec{r_{1}} \rightarrow \vec{r_{2}} \\ \vec{r_{2}} \rightarrow \vec{r_{1}} \end{array} \right) \right)  
\end{eqnarray}
The values of the radii, $b$ and $c$, which depend weakly on the trion momentum $\vec{Q}$ as a result of Pauli blocking, can be varied to minimize the energy. If $b>>c$ or $b<<c$, the trion can be thought of as an electron weakly bound to an exciton. On the other hand, if $b\approx c$, the trion ought to be considered a strongly bonded triplet of two electrons and a hole.

\subsection{Trion Optical Conductivity}
We assume linearly polarized light of frequency $\omega$ and intensity $I_{o}$ incident normally on the $MX_{2}$ monolayer. We assume the same interaction Hamiltonian as given in Eq.~(\ref{eq:light_H}) for the electrons in the valley $\tau=1$. The rate $R_{tr}$ (units: s$^{-1}$-cm$^{2}$) at which trions are generated by the absorption of light can be found using Fermi's golden rule assuming a finite trion lifetime,
\begin{eqnarray}
R_{tr} & = & 2 \times \frac{2\pi}{\hbar} \frac{1}{A} \sum_{\vec{Q}} \left\{ \left| \langle \psi_{tr}(\vec{Q}) | H_{+} | \psi_{i}(\vec{Q}) \rangle \right|^{2} \right\}_{th}\nonumber \\
& & \times \frac{\Gamma_{tr}/\pi}{\left( E_{tr}(\vec{Q}) - E_{i} - \hbar \omega \right)^{2} + \Gamma_{tr}^{2}} \nonumber \\
& = & 2 \times \frac{2\pi}{\hbar} \left( \frac{eA_{o}}{2m_{o}} \right)^{2}  \int \frac{d^{2}\vec{Q}}{(2\pi)^{2}} f_{c}(\vec{Q}) \times  \nonumber \\
& & \left| \int d^{2}\vec{r}_{1} \chi_{tr}(\vec{r}_{1}, \vec{r}_{2}=0, \vec{Q}) e^{-i(m_{ex}/m_{tr})\vec{Q}.\vec{r}_{1}}   \right|^{2} \nonumber \\
& & \times \frac{\Gamma_{tr}/\pi}{\left( E_{tr}(\vec{Q}) - E_{i} - \hbar \omega \right)^{2} + \Gamma_{tr}^{2}}
\end{eqnarray}
Here, the factor of two in the front accounts for the fact that the additional electron binding with the photogenerated electron-hole pair to form a trion can belong to any one of the two valleys, and,
\begin{eqnarray}
& & \chi_{tr}(\vec{r_{1}},\vec{r_{2}},\vec{Q}) =  \int \frac{d^{2}\vec{k_{1}}}{(2\pi)^{2}} \int \frac{d^{2}\vec{k_{2}}}{(2\pi)^{2}} \vec{P}_{cv}(\ul{\vec{k_{2}}}).\hat{n} \, \times \nonumber \\
& & \phi(\vec{k_{1}},\vec{k_{2}},\vec{Q}) \sqrt{1-f_{c}(\ul{\vec{k_{2}}})} \, \, e^{i\vec{k_{1}}.\vec{r_{1}} +i\vec{k_{2}}.\vec{r_{2}} }
\end{eqnarray}
$\chi_{tr}(\vec{r_{1}},\vec{r_{2}},\vec{Q})$ incorporates the reduction in the trion oscillator strength due to Pauli-blocking. Finally, The total energy absorption rate from both $K$ and $K'$ valleys can be written in terms of the trion contribution $\sigma_{tr}(\omega)$ to the optical conductivity,
\begin{eqnarray}
& & \Re\{\sigma_{tr}(\omega)\}  =   \frac{e^{2}}{4\hbar}\left\{ \frac{16\hbar}{m_{o}^{2} \omega} \int \frac{d^{2}\vec{Q}}{(2\pi)^{2}} f_{c}(\vec{Q}) \right. \nonumber \\
& & \left. \times \left| \int d^{2}\vec{r}_{1} \chi_{tr}(\vec{r}_{1}, \vec{r}_{2}=0, \vec{Q}) e^{-i(m_{ex}/m_{tr})\vec{Q}.\vec{r}_{1}}   \right|^{2}  \right. \nonumber \\
& & \left. \times \frac{\Gamma_{tr}}{\left( E_{tr}(\vec{Q}) - E_{i} - \hbar \omega \right)^{2} + \Gamma_{tr}^{2}} \right\} \label{eq:trioncond1}
\end{eqnarray}

\begin{figure}[tbp]
  \begin{center}
   \epsfig{file=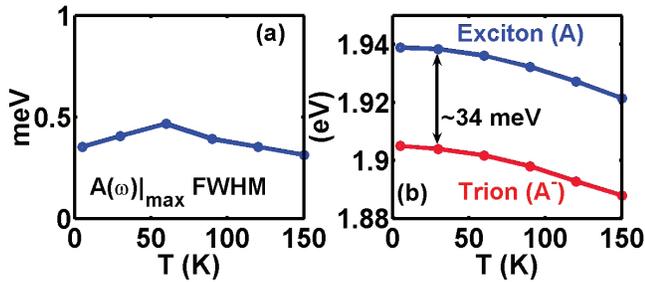,angle=0,width=0.48\textwidth}
    \caption{(a) The product of the peak absorption $A(\omega)|_{max}$ and the FWHM width of the extracted $A^-$-trion absorption spectrum is plotted as a function of temperature. (b) The measured positions of the $A$-exciton peak and the $A^{-}$-trion peak are plotted as a function of temperature. The separation between the peaks does not change much with temperature.}   
    \label{tr_gamma}
  \end{center}
\end{figure} 

\subsection{Experimental Results and Discussion: Trion Absorption in MoS$_{2}$ Monolayers} \label{subsec:trion_abs}
Fig.~\ref{triple_fitting} shows the extracted contributions from $A^-$-trions at 5K and 90K. Fig.~\ref{tr_gamma} shows that the product of the peak absorption $A(\omega)|_{max}$ and the FWHM of the extracted $A^-$-trion absorption spectrum vary little with temperature in the 5K-150K range. The trion absorption spectrum can be related to the trion optical conductivity using Eq.~(\ref{eq:exspectra}). Ignoring Pauli-blocking and using the simple wavevector-independent expression for the momentum matrix element in (\ref{eq:matrix}), an analytical expression for the trion optical conductivity can be found in the limit of small electron density when the Fermi energy $E_{F}$ and the Fermi wavevector $k_{F}$ satisfy $KT<E_{F},\Gamma_{tr}$ and $k_{F}b,k_{F}c <1$,
\begin{eqnarray} 
&& \Re\{\sigma_{tr}(\omega)\} =  \nonumber \\
&& \frac{e^{2}v^{2}}{\omega} \left( \frac{n_{e}}{E_{F}} G(c/b) \right) \frac{m_{tr}}{m_{ex}} \left[ \tan^{-1}\left( \frac{\frac{m_{ex}}{m_{tr}}E_{F}-\Delta E}{\Gamma_{tr}}\right) \right. \nonumber \\
&& \left. + \tan^{-1}\left(\frac{\Delta E}{\Gamma_{tr}}\right) \right] \label{eq:trion_simple}
\end{eqnarray}
Here, $\Delta E = E_{tr}(\vec{Q}=0)-E_{i}-\hbar \omega$. The function $G(x)=G(1/x)$ depends on the ratio of the two radii, $c/b$, and equals,
\begin{equation}
G(x) = 8\left[ 1 + 16 \frac{x^{2}}{(1 + x)^4} \right]^{-1}\left( \frac{1+x^{2}}{x} \right)^{2}
\end{equation}
The above expression shows that when $E_{F}<<\Gamma_{tr}$ the peak trion conductivity increases linearly with the electron density and its spectral shape is almost a Lorentzian that peaks when $\hbar\omega$ equals $E_{tr}(\vec{Q}=0)-E_{i}-0.5 (m_{ex}/m_{tr})E_{F}$ and has a FWHM equal to $2\sqrt{\Gamma_{tr}^{2} + (0.5(m_{ex}/m_{tr})E_{F})^2}$. The expression in Eq.~(\ref{eq:trion_simple}) overestimates the trion conductivity significantly.

\subsubsection{Trion Radii}
For simulations, we use the more accurate expression for the trion conductivity given in (\ref{eq:trioncond1}) that takes into account the decrease in the trion oscillator strength from wavevector-dependent momentum matrix element and Pauli-blocking. The dependence of the trion binding energy $E_{trb}$ on the trion momentum $\vec{Q}$ was ignored. Unlike in the case of excitons, the measured trion absorption spectrum cannot be used to extract both the trion radii, $b$ and $c$, reliably since, as shown in Eq.~\ref{eq:trion_simple}, the trion conductivity is sensitive to the ratio $c/b$. However, if $b$ (or $c$) is known, then the ratio $c/b$ can be determined from the measured absorption spectra. We assume that one of the trion radii, $b$, is approximately equal to the exciton radius that was determined earlier using the measured exciton absorption spectra. This assumption is justified in the next section using the results obtained by solving the trion eigenvalue equation (\ref{eq:eigen2}) and happens to be consistent with the picture of a trion as an electron bound to an exciton. As discussed earlier, the electron density in our samples is in the $2$-$4$$\times10^{12}$ cm$^{-2}$ range. If we assume that $n_{e}=2\times10^{12}$ cm$^{-2}$ then, using Fig.~\ref{ex_radius}(b), we choose $b=9.3\AA$. The resulting value of the ratio $c/b$ that best fits the experimental data in Fig.~\ref{tr_gamma}(a) is found to be $\sim$1.9 ($\Rightarrow c \approx 17.7 \AA$). And if we assume that $n_{e}=4\times10^{12}$ cm$^{-2}$ and $b=8.3\AA$, then $c/b$ is found to be $\sim$1.3. For the smaller electron density assumption, the extracted value ($\sim1.9$) of the ratio $c/b$ is in good agreement with the variational solution of Berkelbach et.~al.~\cite{timothy}, who obtained a value of $\sim$2.4 for $c/b$ but did not include effects due to Pauli-blocking in the trion eigenvalue equation. The extracted value ($\sim1.9$) of the ratio $c/b$ is also in good agreement with results obtained from the trion eigenvalue equation (see the next Section). The good agreement between theory and measurements suggests that the trion oscillator strength, as given by the trion optical conductivity in (\ref{eq:trioncond1}), captures the essential physics. Theoretical predictions for the trion radii that result from the eigenvalue equation (\ref{eq:eigen2}) using the variational solution (\ref{eq:soltrion}) are discussed below.
                
\subsubsection{Trion Binding Energy}
Fig.~\ref{tr_gamma}(b) shows the measured positions of the $A$-exciton peak and the $A^{-}$-trion peak as a function of the temperature. The separation between the peaks is $\sim$34 meV and does not change much with temperature in the 5-150K range. As discussed above, the trion binding energy can be estimated by subtracting $0.5 (m_{ex}/m_{tr})E_{F}$ from the separation between the two peaks. This gives $\sim$32 meV and $\sim$30 meV for the trion binding energy assuming $n_{e}=2\times10^{12}$ cm$^{-2}$ and $n_{e}=4\times10^{12}$ cm$^{-2}$, respectively. These values for the trion binding energy, although larger than the values measured by Fai et.~al.~\cite{fai13}, are in good agreement with the theoretical values reported for MoS$_{2}$ by Berkelbach et.~al.~\cite{timothy}, who obtained $\sim$26 meV and $\sim$32 meV using DFT and GW techniques, respectively. Using a variational solution similar to the one in (\ref{eq:soltrion}), Suris et.~al.~ obtained the result $E_{trb} \approx 0.1 E_{exb}$ when $m_{e} \approx m_{h}$ \cite{suri01}. Using this result, and the exciton binding energies from Fig.~\ref{ex_energy}, the trion binding energy comes out to be $\sim$33 meV and $\sim$28 meV for $n_{e}=2\times10^{12}$ and  $n_{e}=4\times10^{12}$, respectively, again in good agreement with our measurements. The previous theoretical works mentioned here did not take into account Pauli-blocking when solving for the trion wavefunction variationally. Pauli blocking is included in the trion eigenvalue equation (\ref{eq:eigen2}). Solution of the trion eigenvalue equation (\ref{eq:eigen2}) is computationally prohibitive, especially since the binding energy depends on the trion momentum $\vec{Q}$ as a result of Pauli-blocking, and a complete analysis is beyond the scope of this paper. Assuming $n_{e}=2\times10^{12}$, we solved (\ref{eq:eigen2}) using the variational solution in (\ref{eq:soltrion}) for $\vec{Q}=0$. The radii, $b$ and $c$, that minimized the trion binding energy were found to be $9.0\AA$ and $18.9\AA$, respectively, and the ratio $c/b$ came out to be $2.1$ in good (but not perfect) agreement with the value ($\sim$1.9) extracted from trion absorption measurements. The trion binding energy $E_{trb}$ was found to be $\sim$26 meV, which is slightly smaller than the measured value of $\sim$32 meV. For $n_{e}=4\times10^{12}$, the solution of the trion eigenvalue equation gave $b=8.4\AA$, $c=17.6\AA$,  $c/b = 2.09$, and $E_{trb}=22.5$ meV. The results in this Section and in the previous Section show that if the electron density is assumed to have the smallest value within the range of uncertainty in the experimentally measured values then the extracted values of the trion parameters from optical measurements are in better agreement with the theoretical model. Trion optical absorption can potentially be used as a sensitive probe for the carrier density in metal dichalcogenide monolayers. We also point out that since the exact trion binding energy is expected to be larger than the value obtained from a variational solution, the difference between the calculated ($\sim$26 meV) and the measured ($\sim$32 meV) trion binding energies could be due to the inaccuracy of the assumed trion variational solution in (\ref{eq:soltrion}).

\section{Conclusion}
In this work, we presented theoretical models for optical absorption by excitons and trions in 2D metal dichalcogenides. The models presented were based on the Wannier-Mott picture of an exciton and took into account the large spread of the exciton and trion wavefunctions in momentum space as a result of the small exciton and trion radii. Wavevector dependence of the optical matrix elements and phase space blocking due to Pauli exclusion were also incorporated in the models. The experimental optical absorption results for MoS$_{2}$ monolayers showed very good agreements between theory and measurements, and enabled estimations of the exciton and trion radii. The results presented in this paper show that the optical properties (specifically, optical conductivities) of excitons and trions in 2D metal dichalcogenides are adequately described by the models presented here. It should be noted here that unbound electron-hole and exciton-electron scattering states have been ignored in this paper. In general, these correlated states can contribute significantly to the absorption spectrum, especially near the tails of the exciton and trion absorption peaks and near the band edges \cite{runge01}. Our work suggests that near the exciton and trion absorption peaks the contribution from these correlated states is small enough to be ignored. However, more work is needed to fully understand their contribution in strongly interacting dichalcogenide monolayers.

\section{Acknowledgements}
The authors would like to acknowledge helpful discussions with Paul L. McEuen and Michael G. Spencer, and support from CCMR under NSF grant number DMR-1120296, AFOSR-MURI under grant number FA9550-09-1-0705, and ONR under grant number N00014-12-1-0072.

\section{Appendix A: Dielectric Constant}
The dielectric constant $\epsilon(\vec{q},\omega)$ used to obtain the Coulomb potential in Eq .~\ref{eq:eigen} and Eq.~\ref{eq:eigen2} is in general wavevector and frequency dependent. The exciton binding energy determines the frequency of relative motion of the electron and the hole in an exciton. Free carriers, if present, will be effective in screening the Coulomb potential if the exciton binding energy is much smaller than the relevant plasmon energies \cite{excitonbook,kubler}. The measured exciton energies in 2D metal dichalcogenides, and in particular in MoS$_{2}$, are relatively large and in the few tenths of eV range \cite{fai10,fai13}. In electron-doped 2D materials, the plasmon frequency depends on the electron density $n_{e}$ and the wavevector $q$ as $\omega_{p}(\vec{q}) \propto \sqrt{n_{e} \, q}$ \cite{huagbook,mahan}. The relevant wavevectors are those for which $qa < 1$, where $a$ is the exciton radius (see Eq.~\ref{eq:orbital}). Therefore, if the relevant plasmon energy $\hbar \omega_{p}(q=1/a)$ is much smaller than the exciton binding energy then the free electrons will not be able to screen the Coulomb potential effectively. This condition is met for the carrier densities considered in this work. Therefore, it is appropriate to use the unscreened dielectric constant. For large carrier densities, when $n_{e}\ge 0.1/a^{2}$, a screened dielectric constant is a better choice \cite{schmittrink1}. The finite thickness and polarizability of the $MX_2$ monolayer makes the dielectric constant wavevector dependent. $MX_2$ monolayers consist of three atomic layers and the conduction and valence band electrons reside predominantly on the d-orbitals of metal atoms in the center layer. Consider a $MX_2$ monolayer of thickness $d$ and dielectric constant $\epsilon_2$ sandwiched between materials with dielectric constants $\epsilon_1$ and $\epsilon_3$ (See Fig.~\ref{figMX3}). Within the layer of metal atoms the effective dielectric constant is,
\begin{equation}
\epsilon(\vec{q}) = \epsilon_2 \frac{ {1 - \frac{(1-\epsilon_2/\epsilon_1)(1-\epsilon_2/\epsilon_3)}{(1+\epsilon_2/\epsilon_1)(1+\epsilon_2/\epsilon_3)}e^{-2qd}}}{\left[ 1 - \frac{(1-\epsilon_2/\epsilon_1)}{(1+\epsilon_2/\epsilon_1)}e^{-qd} \right] \left[ 1 - \frac{(1-\epsilon_2/\epsilon_3)}{(1+\epsilon_2/\epsilon_3)}e^{-qd} \right]}
\end{equation}
We have used the above expression in our calculations, and this expression also follows from the result obtained by Keldysh~\cite{keldysh79}. In this approximation, the polarizability of the $MX_2$ monolayer is described by an effective dielectric constant $\epsilon_{2}$. If the polarizability of the $MX_{2}$ monolayer is anisotropic then it can be shown that $\epsilon_{2}=\sqrt{\epsilon_{\perp} \epsilon_{\parallel}}$ where $\epsilon_{\perp}$ and $\epsilon_{\parallel}$ are the effective dielectric constants for fields polarized perpendicular and parallel to the plane of the layer, respectively, and the effective layer thickness $d$ entering the above expression equals the actual thickness times a factor equal to $\sqrt{\epsilon_{\parallel} / \epsilon_{\perp}}$. 

\begin{table}[!h]
\begin{ruledtabular}
\centering
\begin{tabular}{|l|l|l|l|} 
\multicolumn{4}{|c|}{Table.1: Dielectric Constant of Bulk MoS$_{2}$}\\ 
\hline
Method & $\epsilon_{\perp}$ &   $\epsilon_{\parallel}$ &  $\sqrt{\epsilon_{\perp}\epsilon_{\parallel}}$ \\
\hline
$LDA$ \cite{molina} & 15.4 & 7.43 & 10.7 \\
\hline
$GW$ \cite{timothy} & 14.29 & 6.87 & 9.9 \\
\hline
$GW$ \cite{Lam12} & 13.5 & 8.5 & 10.7 \\
\hline
$G_{o}W_{o}$ \cite{Rama12} & 14.5 &  &  \\
\end{tabular}
\end{ruledtabular}
\end{table}

For small wavevectors ($q<<1/d$) $\epsilon(\vec{q})$ given above approaches $(\epsilon_1 + \epsilon_3)/2$, and for large wavevectors ($q>>1/d$) $\epsilon(\vec{q})$ approaches $\epsilon_2$. In monolayer $MX_{2}$ materials, since the layer thickness $d$ and the exciton radius $a$ are comparable, the wavevector dependence of the dielectric constant cannot be ignored \cite{timothy}. In our work, $\epsilon_3$ is 4.0 (quartz substrate) and $\epsilon_1$ is 1.0 (free space), and the value of $\epsilon_2$ is used as an adjustable parameter and its value is determined to best fit the measurements (see Section \ref{sec:ex_energy}). Theoretically obtained values for  $\epsilon_2$ for bulk MoS$_{2}$ are presented in the Table.1. Note that in bulk MoS$_{2}$, $\epsilon_{\perp}$ and $\epsilon_{\parallel}$ are different. 

When $qd < 1$, the above expression for $\epsilon(\vec{q})$ becomes,
\begin{equation}
\epsilon(\vec{q}) = \frac{\epsilon_{1} + \epsilon_{3}}{2} \left[ 1 + \frac{2\epsilon_{2}^{2} - (\epsilon_{1}^{2} + \epsilon_{3}^{3})}{2\epsilon_{2}(\epsilon_{1} + \epsilon_{3})} \, q d \right]
\end{equation}
In the limit $qd < 1$, the expression for $\epsilon(\vec{q})$ matches the one derived by Cudazzo et.~al.~\cite{cudazzo11} assuming a strictly two dimensional material (of negligible thickness) and polarizable only in the plane of the material provided the screening length parameter $r_{o}$, which is related to the in-plane polarizability of the two dimensional material, is taken to be,
\begin{equation}
r_{o} =   \frac{2\epsilon_{2}^{2} - (\epsilon_{1}^{2} + \epsilon_{3}^{3})}{2\epsilon_{2}(\epsilon_{1} + \epsilon_{3})} \, d 
\end{equation}
In the case of a MoS$_{2}$ monolayer surrounded by free-space on both sides, the value of $r_{o}$ has been obtained from first principles (DFT-RPA, GW-RPA) by Berkelbach et.~.al~\cite{timothy} and was found to be in the 30-40$\AA$ range. For a MoS$_{2}$ monolayer surrounded by free-space on both sides, the above expression for $r_{o}$ becomes,
\begin{equation}
r_{o} =   \frac{\epsilon_{2}^{2} - 1}{2\epsilon_{2}} \, d 
\end{equation}
Using the value ($\sim$12) of $\epsilon_{2}$ obtained in our work from fitting the optical absorption data (see Section \ref{sec:ex_energy}), and assuming $d\approx 6\AA$, the value of $r_{o}$ using the above expression comes out to be $\sim36\AA$ in excellent agreement with first principles calculations of Berkelbach et.~.al~\cite{timothy}.

\end{document}